# Benchmarking Materials Property Prediction Methods: The Matbench Test Set and Automatminer Reference Algorithm

Alexander Dunn[1,2*], Qi Wang[1], Alex Ganose[1], Daniel Dopp[1], and Anubhav Jain[1*]

1. *Lawrence Berkeley National Laboratory, Energy Technologies Area, 1 Cyclotron Road, Berkeley, CA 94720, United States*

2. *Department of Materials Science and Engineering, University of California, Berkeley CA 94720, United States*

* *Correspondence to ardunn@lbl.gov, ajain@lbl.gov*

## Abstract

We present a benchmark test suite and an automated machine learning procedure for evaluating supervised machine learning (ML) models for predicting properties of inorganic bulk materials. The test suite, Matbench, is a set of 13 ML tasks that range in size from 312 to 132k samples and contain data from 10 density functional theory-derived and experimental sources. Tasks include predicting optical, thermal, electronic, thermodynamic, tensile, and elastic properties given a materials composition and/or crystal structure. The reference algorithm, Automatminer, is a highly-extensible, fully-automated ML pipeline for predicting materials properties from materials primitives (such as composition and crystal structure) without user intervention or hyperparameter tuning. We test Automatminer on the Matbench test suite and compare its predictive power with state-of-the-art crystal graph neural networks and a traditional descriptor-based Random Forest model. We find Automatminer achieves the best performance on 8 of 13 tasks in the benchmark.

We also show our test suite is capable of exposing predictive advantages of each algorithm – namely, that crystal graph methods appear to outperform traditional machine learning methods given ~$10^4$ or greater data points. The pre-processed, ready-to-use Matbench tasks and the Automatminer source code are open source and available online (http://hackingmaterials.lbl.gov/automatminer/). We encourage evaluating new materials ML algorithms on the Matbench benchmark and comparing them against the latest version of Automatminer.

## Introduction

New functional materials are vital for making fundamental advances across scientific domains, including computing and energy conversion. However, most materials are brought to commercialization primarily by direct experimental investigation, an approach typically limited by 20+ year design processes, constraints in the number of chemical systems that can be investigated, and the limits of a particular researcher's intuition. By utilizing materials "big data" and leveraging advances in machine learning (ML), the emerging field of materials informatics has demonstrated massive potential as a catalyst for materials development, alongside *ab initio* techniques such as high-throughput density functional theory[1,2] (DFT). For example, by using support vector machines to search a space of more than 118k candidate crystal structures, Tehrani et al.[3] identified, synthesized, and experimentally validated two novel superhard carbides. In another study, Cooper et al.[4] applied natural language processing (NLP) techniques to assemble 9k photovoltaic candidates from scientific literature; equipped with algorithmic structure-property encodings and a design-to-device data mining workflow, they identified and experimentally realized a new high-performing panchromatic absorption dye. These examples are but two of many. The sheer investigative volume and potential research impact of materials data mining has helped brand it as "materials 4.0"[5] or "the 4th paradigm"[6] of materials research.

However, the growing role of ML in materials design exposes weaknesses in the materials data mining pipeline: first, there is no systematic method for comparing and selecting materials ML models. Comparing newly published models to existing techniques is crucial for rational ML model design and advancement of the field. Other fields of applied ML have seen rapid advancement in recent years in large part due to the creation and use of standardized community benchmarks such *ImageNet*[7] (16,000+ citations) for image classification and the Stanford Question Answering Dataset[8] (1400+ citations) for NLP. While there are commonly used datasets for materials problems as well, *e.g.,* Castelli et al.'s investigation of cubic perovskites[9], it is uncommon for two algorithms to be tested against the same dataset and with the same data cleaning procedures. Methods for estimating generalization error (*e.g.*, the train/test split) also vary significantly. Typically, either the predictive error is averaged over a set of cross-validation folds (CV score)[10] or a hold-out test set is used, with the specifics of the split procedure varying between studies. Furthermore, if a model's hyperparameters are tuned to directly optimize one of these metrics, equivalent to trying many models and only reporting the best one, they may significantly misrepresent the true generalization error[10,11] (*model selection bias*). Arbitrary choice of hold-out set can also bias a comparison in favor of one model over another (*sample selection bias*)[12–14]. Thus, the materials informatics community lacks a *standard benchmarking method* for critically evaluating new models. If models cannot be accurately compared, ML studies are difficult to reproduce and innovation suffers.

Moreover, the breadth of materials ML tasks is so large that many models must still be designed and tuned by hand. While encouraging for the field, the recent explosion[15] of novel descriptors and models has given practitioners a paradox-of-choice, as selecting the optimal descriptors and model for a given task is nontrivial. The consequences of this paradox-of-choice can be that researchers select suboptimal models or spend researcher time towards re-tuning models for new applications. Thus, an automatic algorithm – which requires no expert domain knowledge to operate yet utilizes knowledge from published literature – could be of great use in prototyping, validating, and analyzing novel high-fidelity models.

Given the above considerations, a benchmark consisting of the following two parts is needed: (1) a robust test suite of materials ML tasks and (2) an automatic "reference" model. The test suite must mitigate arbitrarily favoring one model over another. Furthermore, it should contain a variety of datasets such that domain-specific algorithms can compare on specific datasets and general-purpose algorithms can compare across multiple relevant tasks. The second part, the reference algorithm, may serve multiple purposes. First, it might provide a community standard – or "lower bar" – which future innovation in materials ML should aim to surpass. Second, it can act as an entry point into materials informatics for non-domain specialists since it only requires a dataset as input. Finally, it can help determine which descriptors in the literature are most applicable to a given task or set of tasks.

In this paper, we introduce both these developments - a benchmark test set and a reference algorithm - for application to inorganic, solid state materials property prediction tasks. Matbench, the test suite, is a collection of 13 materials science-specific data mining tasks curated to reflect the diversity of modern materials data. Containing both traditional "small" materials datasets of only a few hundred samples and large datasets of $>10^5$ samples from simulation-derived databases, Matbench provides a consistent nested cross validation[16] (NCV) method for estimating regression and classification errors on a range of mechanical, electronic, and thermodynamic material properties. Automatminer, the reference algorithm, is a general-purpose and *fully-automated* machine learning pipeline. In contrast to other published models that are trained to predict a specific property, Automatminer is capable of predicting any materials property given materials primitives (*e.g.*, chemical composition) as input when provided with a suitable training dataset. It does this by performing a procedure similar to a human researcher: by generating descriptors using Matminer's library[17] of published materials-specific featurizations, performing feature reduction and data preprocessing, and determining the best machine learning model by internally testing various possibilities on validation data. We test Automatminer on the test suite in order to establish baseline performance, and we present a comparison of Automatminer with published ML methods. Finally, we demonstrate our benchmark capable of distinguishing predictive strengths and weaknesses among ML techniques. We expect both Matbench and Automatminer to evolve over time, although the current versions of these tools are ready for immediate use. As evidence of its usefulness, Kabiraj et al.[18] have recently used Automatminer in their research on 2D ferromagnets.

# Results

## Matbench test suite v0.1

The Matbench test suite v0.1 contains 13 supervised ML tasks from 10 datasets. Matbench's data is sourced from various sub-disciplines of materials science, such as experimental mechanical properties (alloy strength), computed elastic properties, computed and experimental electronic properties, optical and phonon properties, and thermodynamic stabilities for crystals, 2D materials, and disordered metals. The number of samples in each task ranges from 312 to 132,752, representing both relatively scarce experimental materials properties and comparatively abundant properties such as DFT-GGA[19] formation energies. Each task is a self-contained dataset containing a single material primitive as input (either composition or composition plus crystal structure) and target property as output for each sample. To help enforce homogeneity, datasets are precleaned to remove unphysical computed data and task-irrelevant experimental data (see *Methods* for more details); thus, as opposed to many raw datasets or structured online repositories, Matbench's tasks have already had their data cleaned for input into ML pipelines. We recommend the datasets be used as-is for consistent comparisons between models. To mitigate model and sample selection biases, each task uses a consistent nested cross-validation[16] procedure for error estimation (see

*Methods*). The distribution of datasets with respect to application type, sample count, type of input data, and type of output data is illustrated in Figure 1; detailed notes on each task can be found in Table 1.

**Table 1: The dataset test suite.** The test suite contains 13 separate ML tasks spread across 10 datasets. The test suite's datasets are diversified across multiple metrics, including target property, number of samples (representing several orders of magnitude), and method for determining the target property.

| Target Property (Unit) | Task Type | Data Source | Samples | Structure available | Method |
|---|---|---|---|---|---|
| Bulk Modulus (GPa) | Regression | Materials Project[20–22] | 10,987 | Yes | DFT-GGA |
| Shear Modulus (GPa) | Regression | Materials Project[20–22] | 10,987 | Yes | DFT-GGA |
| Band Gap (eV) | Regression | Materials Project[20,21] | 106,113 | Yes | DFT-GGA |
| Metallicity (binary) | Classification | Materials Project[20,21] | 106,113 | Yes | DFT-GGA |
| Band Gap (eV) | Regression | Zhuo et al.[23] | 4,604 | No | Experiment |
| Metallicity (binary) | Classification | Zhuo et al.[23] | 4,921 | No | Experiment |
| Bulk Metallic Glass formation (binary) | Classification | Landolt-Bornstein Handbook[24,25] | 5,680 | No | Experiment |
| Refractive index (no unit) | Regression | Materials Project[20,21,26] | 4,764 | Yes | DFPT-GGA |

| Formation Energy (eV/atom) | Regression | Materials Project[20,21] | 132,752 | Yes | DFT-GGA |
|---|---|---|---|---|---|
| Formation Energy (eV/atom) | Regression | Castelli et al. [9] | 18,928 | Yes | DFT-GGA |
| Freq. at Last Phonon PhDOS Peak (1/cm) | Regression | Materials Project[20,21,27] | 1,296 | Yes | DFPT-GGA |
| Exfoliation Energy (meV/atom) | Regression | JARVIS DFT 2D[28] | 636 | Yes | DFT-vDW-DF |
| Steel yield strength (GPa) | Regression | Citrine Informatics[29] | 312 | No | Experiment |

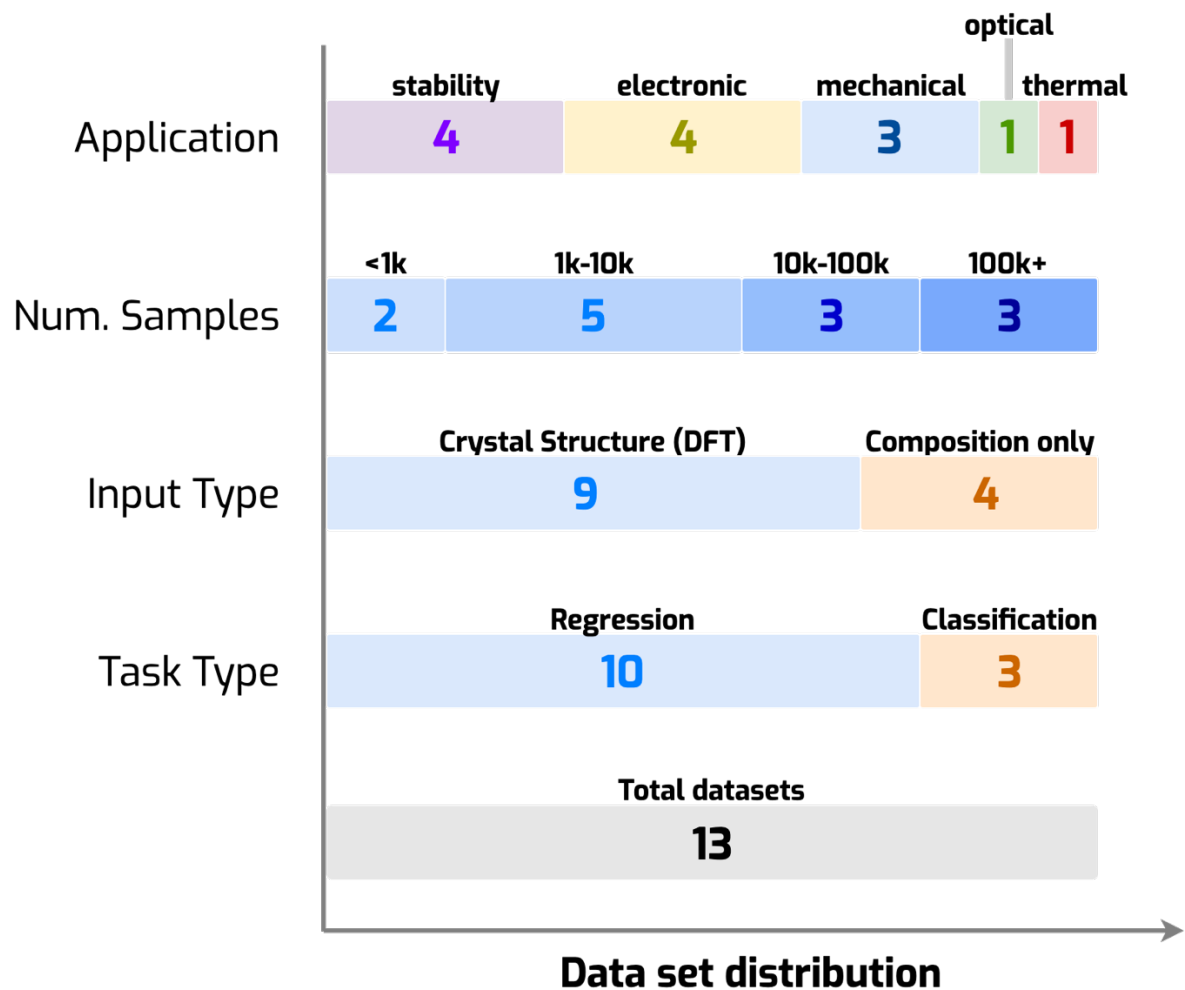

**Figure 1:** Categorical dataset distribution of the 13 machine learning tasks in the Matbench test suite v0.1. Methods of categorization are listed on the left: "Application" describes the ML target property of the task as it relates to materials, "Num. samples" describes the number of samples in each task, "Input Type" describes the materials primitives that serve as input for each task, and "Task Type" designates the supervised ML task type. Numbers in the bars represent the number of tasks fitting the descriptor above it (*e.g.,* there are 10 regression tasks).

## Automatminer Reference Algorithm

At a high level, an Automatminer pipeline can be considered a black box that performs many of the steps typically performed by trained researchers (feature extraction, feature reduction, model selection, hyperparameter tuning). Given only a training dataset, and without further researcher intervention or hyperparameter tuning, Automatminer produces a machine learning model that accepts materials compositions and/or crystal structures and returns predictions. Automatminer can create persistent end-to-end pipelines containing all internal training data, configuration, and the best-found model - allowing the final models to be further inspected, shared, and reproduced.

As shown in Figure 2, the Automatminer pipeline is composed of four stages. In the first stage, autofeaturization, Automatminer generates potentially relevant features using Matminer's featurizer library[17] and verifies that each featurizer is valid for a threshold percentage (default 90%) of materials input objects. An example of an invalid behavior would be trying to apply a featurizer that is not parameterized for noble gases to crystals or compounds containing those elements. Automatminer next applies each featurizer in an error-tolerant fashion, expanding a material primitive into potentially many thousands of features derived from published literature. The next step in the pipeline is the cleaning stage. This prepares the feature matrix for ML by handling errors (e.g., imputing unknown values) and encoding categorical features. The third stage uses one or more dimensionality reduction algorithms (*e.g.,* based on Pearson correlation coefficients[30] or principal component analysis[31]) to reduce the feature vector dimension, removing, for example, redundant or linearly dependent sets of features. Finally, an AutoML stage searches a pre-defined space of internal pipelines which are entirely agnostic to materials inputs. These tree-based internal pipelines as implemented in the TPOT library[32] include techniques for normalization, nonlinear transformations, and ML models with corresponding hyperparameter grids. Each stage can be extensively customized to facilitate end-user needs; for example, pipelines can retain custom features, use single models instead of AutoML, and fine tune feature selection hyperparameters. However, pre-configured pipeline presets are available based on memory, CPU, and time constraints, and no user customization is required to train or predict using materials data when using these presets. In this work, we report results generated using the "Express" preset, which is designed to run with a maximum AutoML training time of 24 hours.

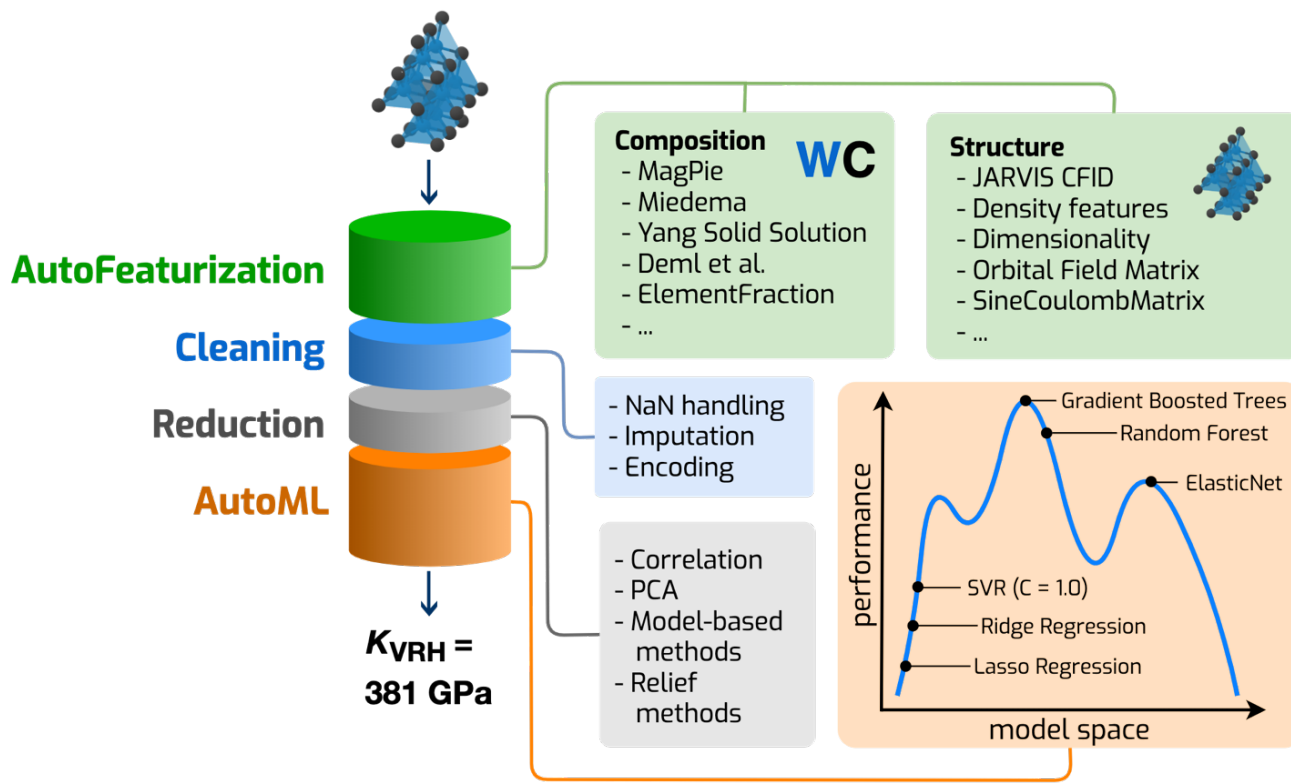

**Figure 2:** The AutoML + Matminer (Automatminer) pipeline, which can be applied to composition-only datasets, structure datasets, and datasets containing electronic bandstructure information. Once fit, the pipeline accepts one or more materials primitives and returns a prediction of a materials property. During autofeaturization, the input dataset is populated with potentially relevant features using the Matminer library. Next, data cleaning and feature reduction stages prepare the feature matrices for input to an AutoML search algorithm. During training, the final stage searches ML pipelines for optimal configurations; during prediction, the best ML pipeline (according to validation score) is used to make predictions.

We evaluate Automatminer on the Matbench test suite and provide comparisons with alternative algorithms in Figure 3. The evaluation is performed using a five-fold Nested Cross Validation (NCV) procedure. In contrast to relying on a single train-test split, in the five-fold NCV procedure, five different train-test sets are created. For each of the five train-test sets, a machine learning model is fit using only the training data and evaluated on the test data. Note that this implies that even for a single type of model (e.g., Automatminer or CGCNN[33]), a slightly different model will be trained for each of the five splits since the training data differs between splits. The errors from the five different overall runs are averaged to give the overall score. Note that within each of the five runs of this outer loop, the training data portion is generally split using an inner cross-validation that is used for model selection within the training data, hence the name "Nested Cross Validation" (in our procedure, an algorithm can make use of the training data however it chooses). One advantage of 5-fold nested CV over a traditional train-test split is that each sample in the overall dataset is present as training in four of the splits and as test in one of the splits.

For all tasks, the Automatminer "Express" preset configuration is used in this work. The Express preset only implements featurizers from Matminer that are broadly applicable (tend to produce valid feature values for almost all compositions and/or crystal structures), are computationally efficient (<2s/sample), and can be trivially transformed from matrices to vectors for each sample. "Express" feature reduction typically retains between 20 and 200 features based on a feature importance threshold from a Random Forest[34] model. The reduced number of features allows for accelerated evolution of the TPOT genetic algorithm within the Express training time limit of 24 hours. Further details can be found in the *Methods* and *Supplementary Information*. While other presets are available in Automatminer, we have found that the Express preset generally retains 95% or more of the accuracy of more expensive presets on multiple data-scarce tasks (bulk metallic glass classification, experimental band gap regression/classification, exfoliation energy regression) at less than 50% of the computational cost to reach reasonable AutoML convergence. We emphasize that the Automatminer Express preset is a *single* configuration capable of fitting on *all* Matbench tasks with *no* additional input or configuration. We do not modify this preset for different tasks.

Four alternative algorithms are used for comparison. To simulate a control, a Dummy model predicts the mean of the training set (regression) or randomly selects a label in proportion to the distribution of the training set (classification). As a second baseline representing commonly used methods, we employ a Random Forest[34] model (RF) using Magpie elemental statistics[25] and Sine Coulomb Matrix[35] (if structures are present in the dataset) to predict each property. Finally, for tasks containing relaxed structures, we also test against CGCNN[33] and MEGNet[36], two graph-network algorithms for general-purpose property prediction. It must be emphasized that a goal of Matbench is to minimize arbitrary biases when comparing models. Therefore, the four alternatives and Automatminer all underwent identical error estimation procedures (NCV on identical folds) for each task.

For some Matbench tasks, we were able to find published scores of researcher-optimized machine learning models, which we label as the "Best Literature" score. However, it should be noted that although these studies report the same error metric (MAE) using *similar* datasets, the scores do not use *identical* datasets (*e.g.*, using different data filtering algorithms to remove erroneous or unreliable data points) or the same error estimation procedure (*e.g.*, they do not use nested cross validation and may use different proportions of train and test). Therefore, these scores cannot be directly compared to the algorithms listed above.

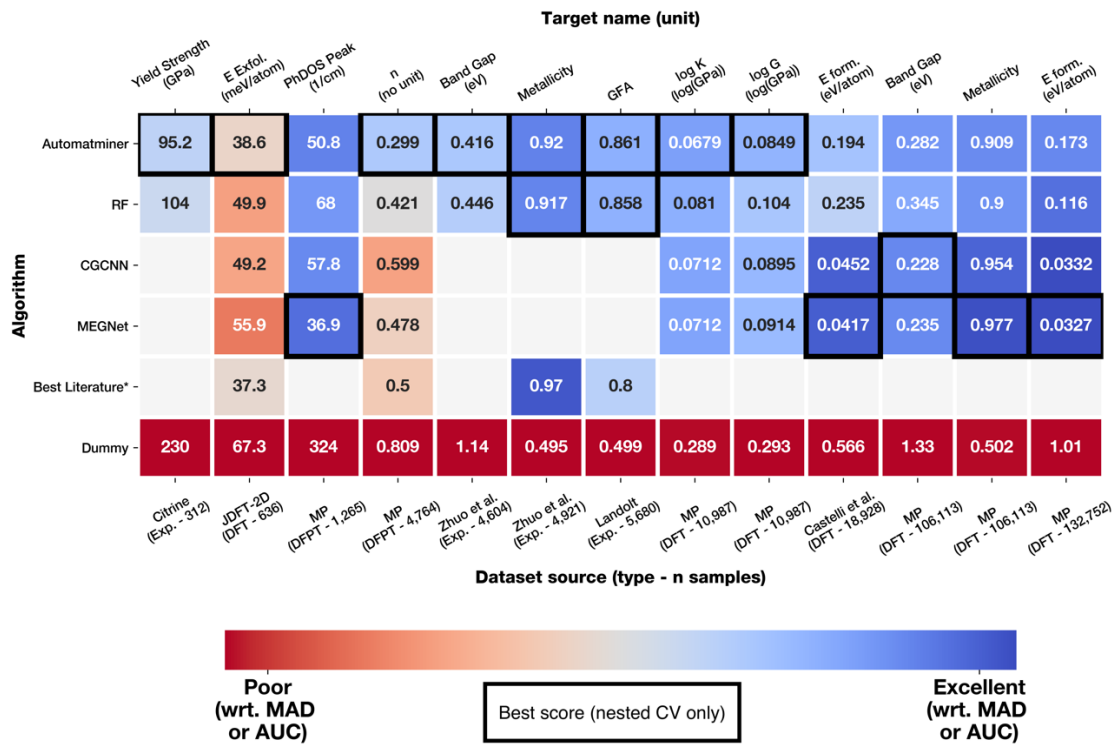

**Figure 3:** Comparison of machine learning algorithm accuracies on the Matbench v0.1 test suite (see Table 1 for more details of the test sets). Numbers on each square represent either the mean average error (regression) or mean ROC-AUC (classification) of a five-fold nested cross validation (NCV), except for "Best Literature" scores. Best Literature scores were taken from published literature models[23,37,38] evaluated on similar tasks or datasets, often subsets of those in Matbench, and do not use NCV. Colors represent "prediction quality" with respect to either the dataset target mean average deviation (MAD) or the high/low limits of ROC-AUC (0.5 is equivalent to random, 1.0 is best); blue and red represent high and low prediction qualities, respectively. The best score for each task is outlined with a black box (The "Best Literature" scores are excluded because they do not use the same testing protocol). To account for variance from choice of NCV split, multiple scores may be outlined if within 1% of the true "best" score. A comparison with a pure Random Forest (RF) model using Magpie[25] and SineCoulombMatrix[35] features is provided for reference. Dummy predictor results are also shown for each task. All Automatminer, CGCNN, MEGNet, and RF results were generated using

the same NCV test procedure on identical train/test folds; all featurizer (descriptor) fitting, hyperparameter optimization, internal validation, and model selection were done on the training set only. A full breakdown of all error estimation procedures can be found in *Methods*.

All models outperform Dummy on all tasks: the Dummy comparison exhibits errors between 68% and 299% higher than the best model for any task. We next examine which algorithms perform best, with "best" taken to include scores within 1% of the best NCV score (we find the standard deviation between folds for the same model is typically between 0.5 - 5%). The Automatminer Express preset has a best NCV score (lowest mean average error, MAE or highest receiver operating characteristic area under curve, ROCAUC) on 8 of 13 tasks. In particular, Automatminer equals or outperforms the RF pipeline on all tasks except predicting formation energies across the Materials Project. Among the nine structure tasks only, Automatminer and MEGNet both have best scores on 4 tasks each. CGCNN is the highest performer only for the Materials Project band gap regression task; yet, across the 6 tasks with more than $10^4$ samples, the MEGNet and CGCNN scores are generally quite close.

Notably, we also find Automatminer has similar errors to scores taken from literature. Although these results are taken directly from published reports which use similar – but not identical – datasets and a variety of non-NCV error procedures, it is notable that Automatminer can *automatically* generate models of roughly similar quality to tediously hand-optimized models. This suggests that similar results as those obtained in the literature can be obtained from a fully automated ML pipeline that requires no researcher tuning or intuition.

Next, we examine how the performance of the various machine learning algorithms varies with the size of the training dataset without regard to the specific task. To do this, we normalize the errors on the various tasks by dividing the mean average error (MAE) by the mean average deviation (MAD) in the dataset. With this normalization, a model that always predicts the average of the dataset will have an error of exactly 1.0. Using least-squares linear regression, we find noticeable inverse trends in the MAE/MAD relative error (Figure 4) with respect to the log of dataset size. Interestingly, irrespective of the target property, the rates of improvement with increasing dataset size (slope of the lines) are vastly different between algorithms. In Figure 4(a), we plot the trend for structure-based regression tasks only. The graph network models CGCNN and MEGNET have high relatively high errors on tasks with small datasets, but improve rapidly as the task's dataset size increases. In contrast, the descriptor-based Automatminer and RF models have lower errors on small datasets, but their rates of improvement are far shallower, and they lose their small data advantage as the data size passes $10^4$ samples. Both graph neural network approaches have similarly high rates of improvement, which may indicate that the underlying ML algorithms are able to leverage information from large datasets more efficiently than traditional ML (RF) or AutoML. This finding corroborates Schmidt et al.'s prediction[15] that universal graph neural networks[33,36] will dominate the state-of-the-art on large (>$10^5$ samples) materials datasets.

In Figure 4(b), we compare Automatminer against the Random Forest model since these two models are able to make predictions on all regression tasks (both composition-only as well as composition plus structure tasks). In 4(b), AutoML's advantage over more conventional techniques narrows as the number of samples increases. Near $10^5$ samples, the AutoML advantage is essentially lost. This phenomenon can be partially explained from the 24-hour training time limitation of the Automatminer Express preset. Although the exact pipeline used by the RF model exists in the Express model space, the long training time of each ML pipeline reduces the AutoML search efficiency. Given enough time and computational resources to internally validate and improve its model, it is highly probable the Automatminer Express preset will either find a model equivalent to or superior to the RF model. However, simple ML models (such as the RF we tested) can equal or outperform our AutoML approach if the AutoML search is inefficient in finding the optimal model.

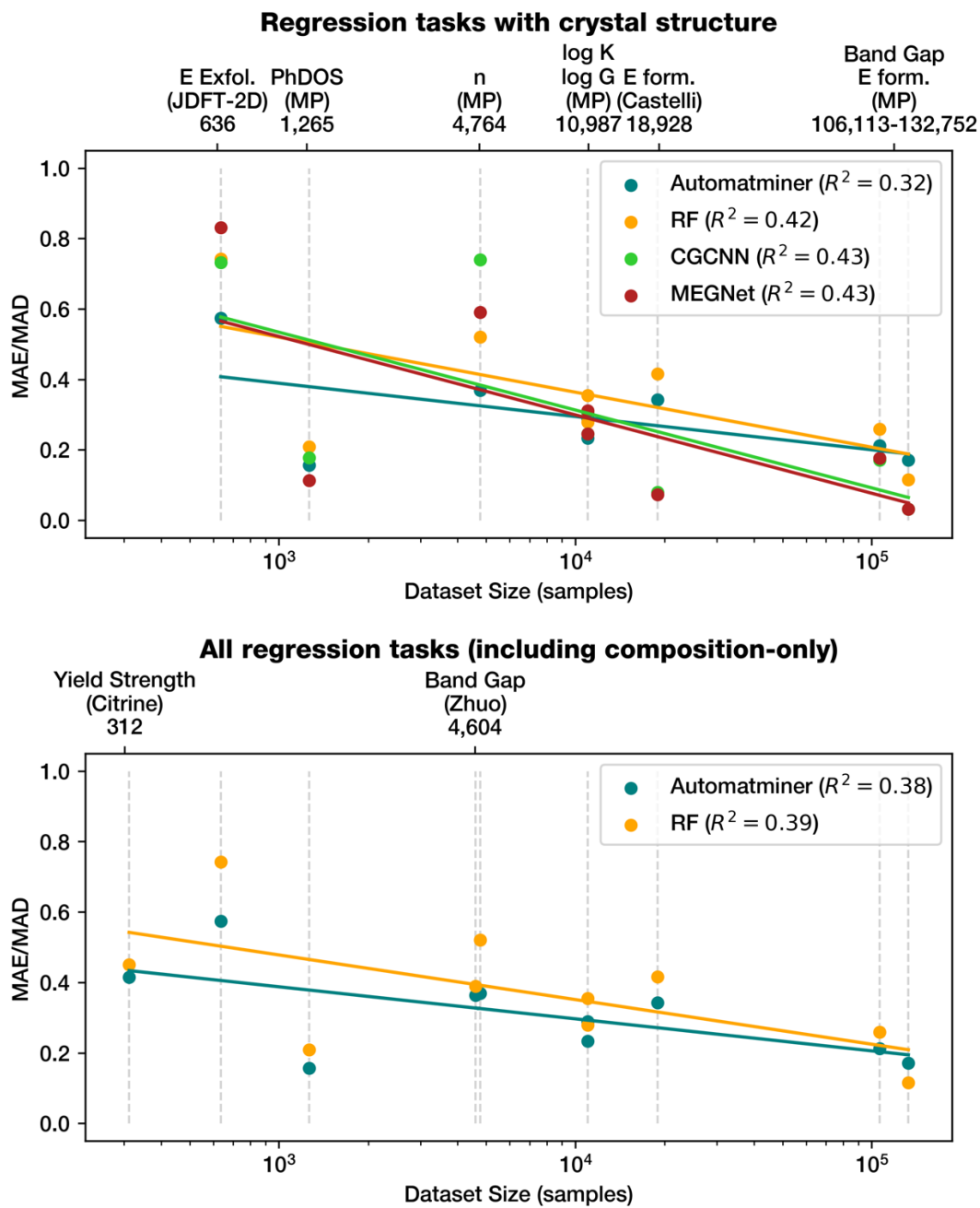

**Figure 4: (a)** Trends in relative predictive accuracy for all algorithms on the eight Matbench v0.1 regression tasks with crystal structure. Algorithms are segregated by color. For each task-algorithm pair, the mean MAE of the nested CV test folds is divided by the dataset mean average deviation to get the relative error. A relative error of zero represents perfect predictive performance; a relative error of 1.0 is equivalent to predicting the mean of the dataset (as in the Dummy Predictor). The plot is agnostic to target property. A least-squares linear regression line of the same color as the scatter points was fit for each algorithm. Multiple tasks have an identical dataset size but differ in their relative errors (*e.g.,* $\log_{10} K$ and $\log_{10} G$). **(b)** Similar to (a) but for all regression tasks (including those lacking crystal structure data as input) and only showing the two algorithms valid for all such tasks.

All algorithms exhibit a noisy yet universal trend which decreases the relative errors as the dataset size increases, even though the underlying task is also changing with size. Such a trend corroborates Zhang and Ling's observations[39] based on a survey of materials ML data in published literature, which suggests the relationship between error (constructed using literature CV data and scaled by range rather than mean average deviation) and dataset size can be fit with a decreasing power law. This trend identified by Zhang and Ling is similar to that found in the more structured results we present. However, we additionally find that the rate of improvement differs substantially between more conventional machine learning approaches versus the graph neural network approaches. Furthermore, despite these overall trends, it is clear that the details of the underlying task do matter. For example, the two graph networks (CGCNN and MEGNET) appear to far outperform the two traditional ML algorithms (Random Forest and Automatminer) on the two formation energy prediction tasks. However, they do not outperform the traditional algorithms by as much on the band gap regression task, despite the large-data domain that graph networks excel in. Similarly, while Automatminer outperforms the graph networks on most small datasets, MEGNet decisively outperforms Automatminer for the PhDOS task. The predictive advantage may lie in MEGNet's specific architecture and implementation rather than an inherent advantage of crystal graph neural networks, given CGCNN has higher error than both Automatminer and MEGNet for the PhDOS task.

# Discussion

The reference algorithm and test suite presented above encompass a benchmark that can be used to accelerate development of supervised learning tasks in materials science. Automatminer provides an extensible and universal platform for automated model selection, while Matbench defines a consistent test procedure for unbiased model comparison. Together, Automatminer + Matbench define a performance baseline for machine learning models aiming to predict materials properties from composition or crystal structure. In this section, we address limitations and extensions of both the reference algorithm and the test suite.

## Reference algorithm analysis

Although the "Express" preset was used to demonstrate Automatminer's performance, the Automatminer pipeline is fully configurable at each stage. To reduce the complexity of developing end-to-end materials ML pipelines, Automatminer provides other preset configurations for varying CPU capabilities, time requirements, and objectives. Each preset defines a specific balance between computational cost and comprehensiveness of ML search. For example, the "Debug" preset employs only a single computationally inexpensive featurizer (Magpie featurizer[17,25]) and a heavily restricted AutoML model space restricted to a two minute training time; similarly, the "Debug_single" preset only uses a single predictor (Random Forest) in place of an AutoML algorithm. Other presets exist which expand on the Express featurizer set using more expensive featurization and longer AutoML optimization times. Generally, we observe diminishing returns on performance with more expensive presets; minor improvements in performance require significant increases in computational time. This is particularly noticeable on small datasets where many ML pipelines can be attempted within the time restriction. For instance, in classifying experimental metallicities, the Express preset improves ROC-AUC a negligible ~0.2% (0.919) on average over Debug (0.917), with the Heavy (most expensive) preset improving only another 0.6%

(0.925). Further details on the comparison of presets can be found in the *Supplementary Information.*

Automatminer may be further improved by including more descriptor techniques in its featurizer sets, especially if those featurizers provide information-dense features at low computational cost. For example, Automatminer does not implement any features for determining $2^{nd}$-nearest neighbor coordination, an important structural motif representing medium-range order. Lack of relevant featurizers may also explain the graph networks' advantages in predicting certain thermodynamic properties. Due to the ability of crystal graph networks to effectively convolve site/bond data, they may more accurately represent 3D chemo-spatial information than traditional descriptors. Future Automatminer development might benefit from using the chemo-spatial data (hidden-layer embeddings) from crystal graph networks as input via transfer learning; similarly, graph-composition networks such as RooSt[40], which have demonstrated success in learning hidden representations from stoichiometry alone, may serve as a valuable improvement on Automatminer's current featurizer set. Adding such descriptors to Automatminer is well within its current capabilities, since Automatminer is extensible (with respect to featurizers) by design.

With respect to machine learning models searched by the AutoML library, we find that the majority of AutoML training on materials ML tasks find tree-ensemble methods perform better than the other models in the search space such as $k$-nearest neighbors, logistic regression, and elastic net regression. On small datasets, we observe tree-ensembles have sufficient model complexity to model material-property relationships more faithfully than regularized linear methods or logistic regression. However, the dominance of tree-ensembles is in part an artifact of the relatively small model search space of Automatminer, which at present does not include nonlinear support vector machine kernels or neural networks. Models with higher complexity, such as deep neural networks, may also improve Automatminer's performance on large datasets. Thus, the AutoML search can be improved by expanding the model space at increasing computational cost. However, regardless of the pre-defined model space or feature set construction, thoughtfully-engineered models such as graph networks or other concepts will likely be able to exceed the baseline AutoML model's performance. An AutoML algorithm is best suited for the *rapid prototyping* of more complex human-tuned models rather than the replacement of architectures designed with human expertise.

## Test suite limitations and extensions

In the Matbench benchmark, we use NCV as a one-size-fits-all tool for evaluation, but it is also conceivable that domain-specific methods better estimate the generalization error than NCV. Ren et al.[38] use "grouped" CV to estimate the error of their models for classifying bulk metallic glasses outside of the chemical systems contained in the training set. The rationale behind grouped CV is that the testing procedure should mimic the real-world application. In the case of bulk metallic glass study, the intended goal of the algorithm was to make predictions in chemical systems where no data points were yet present. However, a randomized train/test split would likely result in selecting *some* data points from *all* chemical systems for the training and testing data. Instead, grouped CV will first separate data points by chemical space, and then select an entire chemical space to fall into either the test or training set. This ensures that testing is conducted on new chemical spaces for which there is no training data within that chemical space.

Yet, using grouped CV requires a well-defined manner for grouping the data. In the case of bulk metallic glasses, chemical systems are easily identified as natural groups since the goal is to predict data for entirely unexplored chemical systems. For other materials ML tasks, features for grouping may be hidden in subtle structural motifs or nuances of electronic configuration. Leave-one-cluster-out CV (LOCO-CV)[41] is one potential variant of grouped CV that aims to automate grouping by $k$-means clustering. However, the groups are determined by the choice of input features, which poses two fundamental problems with this technique. First, researchers employing different input features will end up with different definitions of groups and thus different testing procedures; this could be corrected if the features used for the grouping procedure were standardized (even if a different set of input features was used for prediction). Second, the input features may not properly capture the most physically-relevant grouping; for example, if all input features are based on composition, but the most natural grouping is by a structural feature such as crystal type, then the resulting groups will have less value. Thus, for now it is largely up to researchers to determine the need for using grouped CV and to determine the best grouping strategy. Other strategies[41,42] to predict outlier data in the test set may also prove useful.

An improved benchmark could use a specific, distinct error estimation procedure for every task; such a procedure can be determined by domain experts to most accurately represent the real-world use of the algorithm. The ideal benchmark would therefore be a consensus of community tasks, each with an error estimation procedure customized to most accurately reflect the algorithm's *true* error rate in that particular subfield. We chose NCV as a standard error estimator because there are few such well-agreed-upon procedures for existing materials datasets. Future versions of the benchmark may include error estimation procedures other than NCV.

Matbench is not intended to be a final benchmark but a versioned resource that will grow with the field. The ever-increasing volume of data generated from advances in high-throughput experimentation and computation may enable future ML algorithms to predict classes of materials properties that are presently sparse. For example, *ab initio* defect calculations are presently expensive, but an investigation by Emery and Wolverton[43] has demonstrated DFT can generate defect data in promising quantities for future mainstream statistical learning. Advances in high-throughput experimental techniques (such as automated experimentation) also have the possibility to vastly increase the size and scope of materials data; for instance, a recent study[44] was able to capture UV-Vis spectroscopy data for more than 179,000 metal oxides. A benchmark must evolve to represent these advancements in materials data production. We expect Matbench to be an evolving representation of materials property prediction tasks, and updated versions of Matbench will be released to reflect emerging areas of research. In a similar fashion, Automatminer is designed to be extensible toward new techniques for generating descriptors from compositions, crystal structures, and electronic band structures. As new research is released for converting materials objects to machine-learnable descriptors, we intend on incorporating this knowledge into Automatminer's architecture.

## Conclusion

We presented Matbench v0.1, a set of ML tasks aimed at standardizing comparisons of materials property prediction algorithms. We also introduced Automatminer, a fully-automated pipeline for predicting materials properties, which we used to set a baseline across the task set. Using Matbench, we compare Automatminer with crystal graph neural network models, a traditional Random Forest model, and a Dummy control model. We find Automatminer's auto-generated models outperform or equal the RF model across all but one task and are more accurate than crystal graph networks on most tasks with ~$10^4$ points or fewer. However, crystal graph networks appear to learn better on tasks with larger datasets. Automatminer can be used outside of benchmarking to make predictions automatically and seed research for more specialized, hand-tuned models. We encourage evaluating new ML algorithms on the Matbench benchmark and comparing with the latest version of Automatminer.

## Methods

Raw data for Matbench v0.1 were obtained by downloading from the original sources. Tabular versions of some datasets are available online through Matminer's dataset retrieval tools. These datasets contain metadata and auxiliary data. In contrast, the final Matbench datasets are curated tasks containing only the materials input objects and target variables, with all extraneous data removed.

Unphysical (*e.g.,* negative DFT elastic moduli), highly uncommon or unrepresentative samples (*e.g.*, solid state noble gases) were removed according to a specific per-task procedure. Table 2 describes the resources and steps needed to recreate each dataset from the original source or Matminer version.

**Table 2: Procedures and sources for creating datasets in Matbench v0.1.**
"Original Source" denotes the original work that produced the raw data, which needs not be in tabular form. Matminer source datasets are tabular versions of this raw data which can be retrieved with Matminer and may apply additional post-processing or filtering to the original source data. More information on these datasets can be found on Matminer's dataset summary page and in the Matminer source code. Additional modifications are enumerated.

| Task name | Target Property (Unit) | Original Source | Matminer source dataset | Additional modifications |
|---|---|---|---|---|
| log_kvrh | Bulk Modulus (GPa) | Materials Project[20–22] | None* | 1,2,3,6,7 |
| log_gvrh | Shear Modulus (GPa) | Materials Project[20–22] | None* | 1,2,3,6,7 |
| mp_gap | Band Gap (eV) | Materials Project[20,21] | None* | 1,6,7 |
| mp_is_metal | Metallicity (binary) | Materials Project[20,21] | None* | 1,6,7 |
| expt_gap | Band Gap (eV) | Zhuo et al.[23] | expt_gap | 8,9 |
| expt_is_metal | Metallicity (binary) | Zhuo et al.[23] | expt_gap | 8,10 |
| glass | Bulk Metallic Glass formation (binary) | Landolt-Bornstein Handbook[24,25] | glass_ternary_landolt | 8, 11 |
| dielectric | Refractive index (no unit) | Materials Project[20,21,26] | None* | 1,4,6,7 |
| mp_e_form | Formation Energy (eV/atom) | Materials Project[20,21] | None* | 5,6,7 |

| perovskites | Formation Energy (eV/atom) | Castelli et al.[9] | castelli_perovskites | 7 |
|---|---|---|---|---|
| phonons | Freq. at Last Phonon PhDOS Peak (1/cm) | Materials Project[20,21,27] | phonon_dielectric_mp | 1,7 |
| jdft2d | Exfoliation Energy (meV/atom) | JARVIS DFT 2D[28] | jarvis_dft_2d | 7 |
| steels | Steel yield strength (GPa) | Citrine Informatics[29] | steel_strength | 8 |

\* Generated using the Materials Project API[21] on 4/12/2019.

1. Remove entries having a formation energy or energy above the convex hull more than 150meV.
2. Remove entries having $G_{\text{Voigt}}$, $G_{\text{Reuss}}$, $G_{\text{VRH}}$, $K_{\text{Voigt}}$, $K_{\text{Reuss}}$, or $K_{\text{VRH}}$ less than or equal to zero.
3. Remove entries failing $G_{\text{Reuss}} \leq G_{\text{VRH}} \leq G_{\text{Voigt}}$ or $K_{\text{Reuss}} \leq K_{\text{VRH}} \leq K_{\text{Voigt}}$
4. Remove entry with refractive index less than 1.
5. Remove entries having formation energies greater than 3.0eV. This operation removes ~1500 1-dimensional crystal structures likely resulting from mis-converged DFT structure optimizations of Half-Heuslers present in the Materials Project database as of the generation date.
6. Remove entries containing noble gases.
7. Remove all columns except structure and the target variable.
8. Remove all columns except composition and the target variable.
9. Filter according to unique compositions by ensuring no composition has conflicting metallicity.
10. Filter according to unique compositions by removing compositions with a range of reported band gap values of more than 0.1eV. For each remaining composition, select the value closest to the mean of that composition's reported values.
11. Filter according to unique compositions, removing compositions with any conflicting bulk metallic glass formation classifications.

Five-fold nested cross validation was used to evaluate each algorithm on every task of the benchmark. The outer test loop of the cross validation used uniformly randomized splits generated with scikit-learn[45] KFold (random seed 18012019). The splits were identical for each algorithm. Classification tasks used stratified cross validation generated with StratifiedKFold (random seed 18012019) to more accurately represent classification performance with unbalanced numbers of each class label. Within each of the five splits, 80% training + validation data is given to the algorithm to optimize the model internally, and the remaining 20% is used for testing. After predicting on each of the five 20% test splits, the error or AUC is averaged over the five folds. The internal validation and model selection process is dependent on the algorithm.

It is worthwhile to quickly enumerate the limitations of NCV and justify its use. First, NCV is computationally expensive. For $k$-fold NCV, the traditional hold-out tuning/validation/test procedure must be repeated $k$ times. NCV also depends on the choice of internal learning procedure for each fold, an aspect which mimics the selection process used by other resampling methods; thus, even when the test sets are fixed, repeating identical procedures can produce error estimates with high variance[46]. Several alternative schemes have been proposed which preserve NCV's advantages while attempting to mitigate issues from increased variability and computational cost. One potential improvement is repeated NCV; but even this approach demonstrates large variation of loss estimates across nested CV runs and is even more computationally expensive than NCV[47]. A promising alternative proposes a smooth analytical alternative to NCV which would reduce the NCV's computational intensity[46]. This analytical alternative also reduces the variability introduced by learning set choice using weights determined after the outer CV loop has been fixed. Yet, the analytical alternative relies on critical assumptions which do not hold for particular models such as support vector machines with noisy observations. Therefore, at this time, NCV is an adequate method for evaluating and comparing models using the Matbench benchmark.

The descriptor-based RF and Automatminer models use Matminer[17] to generate all descriptors and have identical data cleaning procedures. The Random Forest model uses the SineCoulombMatrix[35] featurizer for tasks containing structure and mean, average deviation, range, and max/min statistics on elemental Magpie[25] features (implemented as the "magpie" preset for the ElementProperty featurizer) for all tasks containing chemical compositions. To handle missing features, the RF pipeline drops features with more than 1% missing values. Remaining samples having missing features are imputed using the mean of the known data. Categorical features were encoded using one-hot encoding. The Random Forest model itself consisted of 500 estimators and a max depth of "None", meaning nodes are expanded until all leaves are pure or contain less than 2 samples.

Automatminer v1.0.3.20191111 was used for all Automatminer benchmarks. Features were generated according to Automatminer's autofeaturizer "Express" preset, and a full list of featurizers is available in the *Supplementary Information*. The number of features was reduced using an ensemble-based decision tree method set to capture 99% of the Gini importance[48]. Finally, TPOT v.10.1 was used to train and internally validate (5-fold CV within the training data) competing ML pipelines before selecting the model used to make test predictions. TPOT uses an evolutionary algorithm to optimize the hyperparameters in a given model space. In this context, algorithms (e.g., support vector machines, gradient boosted trees) are integrated into their existing hyperparameter grids such that the algorithms are treated essentially as special hyperparameters. Internal TPOT pre- and post-processing steps (such as normalization) are also included in the model space. Rather than determining a set number of generations to evolve the model population, the Automatminer "Express" preset sets TPOT to evaluate the maximum number of generations of 100 individuals each within 24 hours given a maximum evaluation time of 20 minutes per individual. Individuals were trained and evaluated with 10x parallelism using the *n_jobs* Automatminer preset configuration option. A full table of the Automatminer-TPOT model space is described in the *Supplementary Information*.

CGCNN and MEGNet models were trained and optimized by splitting the training portion of each outer NCV fold into 75% train and 25% validation portions. Thus, the overall split for each fold is 60% training, 20% validation, and 20% test. Each model is trained in epochs of 128-structure batches by optimizing according to mean squared error loss (regression) or binary cross-entropy (classification). After each epoch, the validation loss is computed with the same scoring functions as the final evaluation: MAE for regression or ROC-AUC for classification (made negative so that higher loss represents worse performance). To prevent overfitting, the training is stopped early when the validation loss does not improve over a period of at least 500 epochs. A full table of hyperparameters for each algorithm is provided in the *Supplementary Information.*

Each model's training, validation, and evaluation for each NCV fold were performed on separate groups of compute nodes. Each fold of the RF model and Automatminer were trained and evaluated on a single 24-core Intel Xeon E5-2670 v3 with 64GB RAM (LR4 node). All CGCNN and MEGNET training was performed using one NVIDIA 1080Ti GPU using CUDA (accompanied by two Intel Xeon E5-2623 CPUs with 60GB RAM). Workflows were set up and executed using the FireWorks[49] software package.

# Data Availability

Instructions for downloading and using the Matbench benchmark can be viewed on the official documentation (https://hackingmaterials.lbl.gov/automatminer/datasets.html). The datasets can also be interactively viewed and examined on the Materials Project MPContribs-ML platform (https://ml.materialsproject.org) as serialized tabular data. The code for retrieving and loading the Matbench datasets can be found in the dataset_retrieval folder of the Matminer code (https://github.com/hackingmaterials/matminer). We also encourage readers to suggest modifications to the Matbench dataset test suite on the help forum (https://discuss.matsci.org/c/matminer).  All versions of the Automatminer code are open source via a BSD-style license and are available through the online repository (https://github.com/hackingmaterials/automatminer). We note that all the code for running the specific tests in this paper is also present in a subpackage of this repository: (https://github.com/hackingmaterials/automatminer/tree/master/automatminer_dev ).

## Author Contributions

AD, QW, and AJ conceived the project. AD, QW, DD, and AJ designed the dataset test suite. AD, QW, and AG implemented the Automatminer codebase. AD and QW performed the benchmarking tests. AD prepared the manuscript. AJ supervised the project. All authors reviewed and edited the manuscript.

## Competing Interests

The authors declare no competing interests.

## Acknowledgements

This work was intellectually led and funded by the United States Department of Energy, Office of Basic Energy Sciences, Early Career Research Program, which provided funding for AD, QW, AG, DD, and AJ. Lawrence Berkeley National Laboratory is funded by the DOE under award DE-AC02-05CH11231. This research used the Lawrencium computational cluster resource provided by the IT Division at the Lawrence Berkeley National Laboratory (Supported by the Director, Office of Science, Office of Basic Energy Sciences, of the U.S. Department of Energy under Contract No. DE-AC02-05CH11231). This research used resources of the National Energy Research Scientific Computing Center (NERSC), a U.S. Department of Energy Office of Science User Facility operated under Contract No. DE-AC02-05CH11231. We thank Samy Cherfaoui of the Electrical Engineering and Computer Science Department at the University of California, Berkeley for code contributions. We also thank Patrick Huck for assistance in hosting the data on the MPContribs platform through the Materials Project.

# Supplementary Information

For *Benchmarking Materials Property Prediction Methods: The Matbench Test Set and Automatminer Reference Algorithm*

By Alexander Dunn, Qi Wang, Alex Ganose, Daniel Dopp, and Anubhav Jain

# 1. Training and prediction timing

**Training Times (All k-NCV Folds)**

Legend:
- Automatminer (CPU) ($y = 0.43x - 0.73, R^2 = 0.53$)
- Random Forest (CPU) ($y = 1.10x - 5.83, R^2 = 0.79$)
- MEGNet (GPU) ($y = 1.20x - 3.98, R^2 = 0.93$)
- CGCNN (GPU) ($y = 1.00x - 3.39, R^2 = 0.98$)

Y-axis: Node Hours

X-axis: Dataset Size (samples)

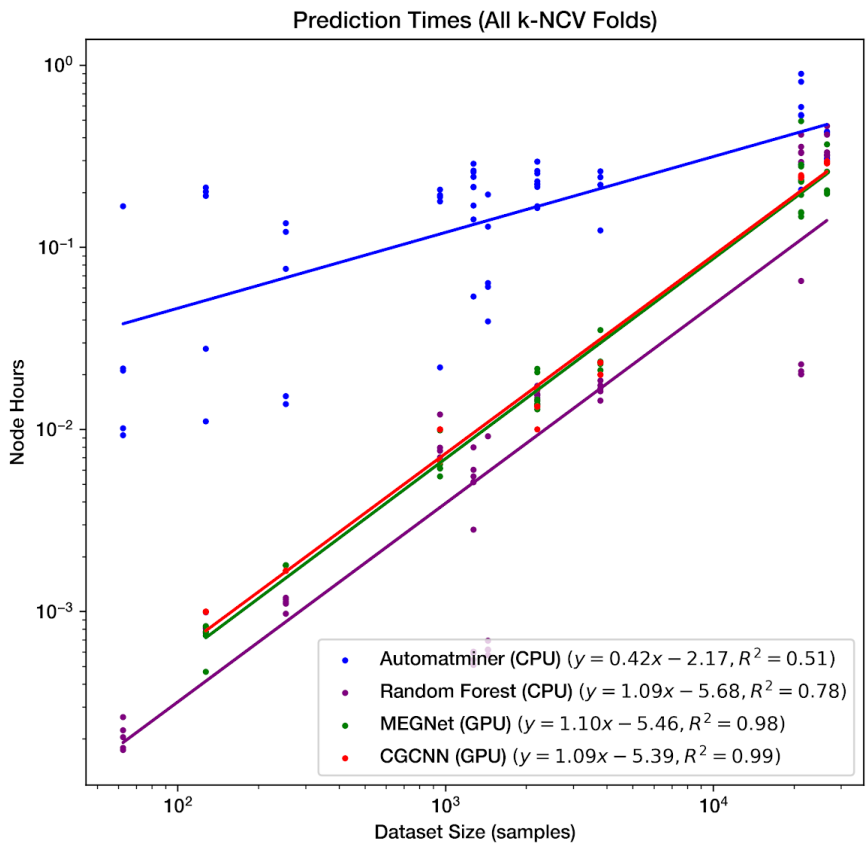

**Figure S1:** Training and prediction times (in node hours) for MEGNet[1], CGCNN[2], Automatminer, and the Random Forest[3] (RF) algorithm on the Matbench test suite in log-log scale. All 13 tasks in Matbench (including composition-only tasks) are shown for Automatminer and the RF algorithms; only the 9 structure tasks are shown for the graph networks MEGNet and CGCNN. Each point represents a single fold of nested cross validation for each task. Each fold of the RF model and Automatminer were trained and evaluated on a compute node containing a single 24-core Intel Xeon E5-2670 v3 with 64GB RAM (LR4 node). Similarly, all CGCNN and MEGNet training and prediction was performed using a single node containing two Intel Xeon E5-2623 CPUs with 60GB RAM) and one NVIDIA 1080Ti (parallelized with CUDA); multiple GPUs were not used as, at the time of training, neither CGCNN nor MEGNet support multi-GPU training. Linear regressions in log-log scale are provided for each algorithm.

# 2. Automatminer Configuration

Beyond what is listed in *Methods,* the Automatminer configuration is determined by the specifics of two primary operations - specifying a set of Matminer[4] featurizers (featurization) and specifying a predetermined model space. During fitting and prediction, Automatminer robustly applies this set of featurizers and TPOT[5] searches the model space for the optimal model. The Automatminer "Express" preset used to generate the results in the main text includes preset settings for a set of Automatminer features (Table S2.1), a data cleaning and feature reduction procedure (described in *Methods*), and settings for TPOT, including the model space (Tables S2.2-3). Alternate presets available in Automatminer can be explored in the source code.

**Table S2.1 Automatminer "Express" preset featurizers.** The following set of Matminer featurizers was used for all Automatminer results in the main text.

| Featurizer Name | Featurizer type | Description |
|---|---|---|
| AtomicOrbitals | composition | HOMO/LUMO orbitals estimated from atomic orbital energies of the composition[6] |
| ElementProperty ("matminer" preset) | composition | Weighted elemental statistics from pymatgen properties[7] |
| ElementProperty ("deml" preset) | composition | Weighted elemental statistics from Deml et al. properties[8] |
| ElementProperty ("matscholar" preset) | composition | Weighted elemental embeddings from Tshitoyan et al. Word2Vec algorithm[9] |
| ElementProperty ("magpie" preset) | composition | Weighted elemental statistics from Ward et al.[10] |
| ElementFraction | composition | Fractions of elements in a composition |
| Stoichiometry | composition | $L_p$ ($0 \leq p \leq 10$) norms of stoichiometric attributes based |

| | | on Ward et al.[10] |
|---|---|---|
| TMetalFraction | composition | Stoichiometric fraction of magnetic transition metal in a composition |
| BandCenter | composition | Electronegativity estimate of absolute band center position using method from Butler and Ginley[11] |
| DensityFeatures | structure | Density, volume per atom, and packing fraction |
| GlobalSymmetryFeatures | structure | Spacegroup and crystal system determination |
| EwaldEnergy | structure | Energy computed from Coulomb interactions using method from Ewald[12] |
| SineCoulombMatrix | structure | A Coulomb matrix[13] variant developed for periodic systems using method from Faber et al.[14] and vectorized using eigenvalues |

**Table S2.2 Automatminer regression model space.** The following table contains the list of classes (containing a scikit-learn BaseEstimator API[15,16]) used in the Automatminer regression model space. Note that many models included in the table are typical preprocessing steps rather than machine learning models; TPOT is capable of stacking these preprocessing steps and ML models into pipelines in a loosely structured tree hierarchy. Hyperparameter grids are defined using the arguments in the "Variable hyperparameter" column. Ranges for each hyperparameter can be found in the Automatminer source code.

| **BaseEstimator (model)** | **Variable hyperparameters** |
|---|---|
| `sklearn.linear_model.ElasticNetCV` | `l1_ratio, tol` |
| `sklearn.ensemble.ExtraTreesRegressor` | `n_estimators, max_features, min_samples_split, min_samples_leaf, bootstrap` |
| sklearn.ensemble.GradientBoostingRegressor | n_estimators, loss, learning_rate, max_depth, min_samples_split, min_samples_leaf, subsample, max_features, alpha |
| sklearn.tree.DecisionTreeRegressor | max_depth, min_samples_split, min_samples_leaf |
| sklearn.neighbors.KNeighborsRegressor | n_neighbors, weights, p |
| sklearn.linear_model.LassoLarsCV | normalize |
| sklearn.svm.LinearSVR | loss, dual, tol, C, epsilon |
| sklearn.ensemble.RandomForestRegressor | n_estimators, max_features, min_samples_split, min_samples_leaf, bootstrap |
| sklearn.linear_model.RidgeCV | None |
| xgboost.XGBRegressor | n_estimators, max_depth, learning_rate, subsample, min_child_weight, nthread |
| sklearn.preprocessing.Binarizer | threshold |
| sklearn.decomposition.FastICA | tol |
| sklearn.cluster.FeatureAgglomeration | linkage, affinity |
| sklearn.preprocessing.MaxAbsScaler | None |
| sklearn.preprocessing.MinMaxScaler | None |
| sklearn.preprocessing.Normalizer | norm |
| sklearn.kernel_approximation.Nystroem | kernel, gamma, n_components |
| sklearn.decomposition.PCA | svd_solver, iterated_power |
| sklearn.preprocessing.PolynomialFeatures | degree, include_bias, interaction_only |
| sklearn.kernel_approximation.RBFSampler | gamma |
| sklearn.preprocessing.RobustScaler | None |
| sklearn.preprocessing.StandardScaler | None |
| tpot.builtins.ZeroCount | None |
| tpot.builtins.OneHotEncoder | minimum_fraction, sparse, threshold |
| sklearn.feature_selection.SelectFwe | alpha, score_func |
| sklearn.feature_selection.SelectPercentile | percentile, score_func |
| sklearn.feature_selection.VarianceThreshold | threshold |

| sklearn.feature_selection.SelectFromModel | threshold, estimator |
|---|---|

**Table S2.3 Automatminer classification model space.** The following table contains the list of classes (containing a scikit-learn BaseEstimator API[15,16]) used in the Automatminer classification model space. Hyperparameter grids are defined using the arguments in the "Variable hyperparameter" column. Ranges for each hyperparameter can be found in the Automatminer source code.

| BaseEstimator (model) | Variable hyperparameters |
|---|---|
| sklearn.naive_bayes.GaussianNB | None |
| sklearn.naive_bayes.BernoulliNB | alpha, fit_prior |
| sklearn.naive_bayes.MultinomialNB | alpha, fit_prior |
| sklearn.tree.DecisionTreeClassifier | criterion, max_depth, min_samples_split, min_samples_leaf |
| sklearn.ensemble.ExtraTreesClassifier | n_estimators, criterion, max_features, min_samples_split, min_samples_leaf, bootstrap |
| sklearn.ensemble.RandomForestClassifier | n_estimators, criterion, max_features, min_samples_split, min_samples_leaf, bootstrap |
| sklearn.ensemble.GradientBoostingClassifier | n_estimators, learning_rate, max_depth, min_samples_split, min_samples_leaf, subsample, max_features |
| sklearn.neighbors.KNeighborsClassifier | n_neighbors, weights, p |
| sklearn.svm.LinearSVC | penalty, loss, dual, tol, C |
| sklearn.linear_model.LogisticRegression | penalty, C, dual |
| xgboost.XGBClassifier | n_estimators, max_depth, learning_rate, subsample, min_child_weight, nthread |
| sklearn.preprocessing.Binarizer | threshold |
| sklearn.decomposition.FastICA | tol |
| sklearn.cluster.FeatureAgglomeration | linkage, affinity |
| sklearn.preprocessing.MaxAbsScaler | None |
| sklearn.preprocessing.MinMaxScaler | None |
| sklearn.preprocessing.Normalizer | norm |
| `sklearn.kernel_approximation.Nystroem` | `kernel, gamma, n_components` |
| `sklearn.decomposition.PCA` | `svd_solver, iterated_power` |
| `sklearn.preprocessing.PolynomialFeatures` | `degree, include_bias, interaction_only` |
| `sklearn.kernel_approximation.RBFSampler` | `gamma` |
| `sklearn.preprocessing.RobustScaler` | `None` |
| `sklearn.preprocessing.StandardScaler` | `None` |
| `tpot.builtins.ZeroCount` | `None` |
| `tpot.builtins.OneHotEncoder` | `minimum_fraction, sparse, threshold` |
| `sklearn.feature_selection.SelectFwe` | `alpha, score_func` |
| `sklearn.feature_selection.SelectPercentile` | `percentile, score_func` |
| `sklearn.feature_selection.VarianceThreshold` | `threshold` |
| `sklearn.feature_selection.RFE` | `step, estimator` |
| `sklearn.feature_selection.SelectFromModel` | `threshold, estimator` |

# 3. Convolutional Graph Network Hyperparameters

**Table S3.1. CGCNN hyperparameters.** The training of CGCNN models is based on its official repo ([https://github.com/txie-93/cgcnn](https://github.com/txie-93/cgcnn)). The hyperparameters are recommended in the CGCNN paper and are used by the authors to generate the pretrained models placed in the repo. Identical hyperparameters were used to generate the results for all tasks in the test suite.

| CGCNN Hyperparameters | Setting |
|---|---|
| Convolution layers | 4 |
| Epochs | 1000 (regression) / 500 (classification) |
| Initial atomic feature length | 92 |
| Atomic feature length | 64 |
| Bond feature length | 41 |
| Hidden feature length | 32 |
| Batch size | 128 |
| Optimization algorithm | Stochastic gradient descent |
| L2 hidden layers | 1 |
| Learning rate | 0.02 |
| Learning rate milestones | 800 |
| momentum | 0.9 |

**Table S3.2 CGCNN full model breakdown.** The number of hyperparameters in a neural network in relation to the number of training samples can be considered a useful metric for analyzing the networks risk of overfitting. The table contains a full breakdown of the model including the number of hyperparameters for each network level. In total, there are 98,185 trainable parameters.

| Layer | Number of parameters |
|---|---|
| embedding.weight | 5,888 |
| embedding.bias | 64 |
| convs.0.fc_full.weight | 21,632 |
| convs.0.fc_full.bias | 128 |
| convs.0.bn1.weight | 128 |
| convs.0.bn1.bias | 128 |
| `convs.0.bn1.running_mean` | 128 |
| `convs.0.bn1.running_var` | 128 |
| `convs.0.bn1.num_batches_tracked` | 1 |
| `convs.0.bn2.weight` | 64 |
| `convs.0.bn2.bias` | 64 |
| `convs.0.bn2.running_mean` | 64 |
| `convs.0.bn2.running_var` | 64 |
| `convs.0.bn2.num_batches_tracked` | 1 |
| `convs.1.fc_full.weight` | 21,632 |
| `convs.1.fc_full.bias` | 128 |
| `convs.1.bn1.weight` | 128 |
| `convs.1.bn1.bias` | 128 |
| `convs.1.bn1.running_mean` | 128 |
| `convs.1.bn1.running_var` | 128 |
| `convs.1.bn1.num_batches_tracked` | 1 |
| `convs.1.bn2.weight` | 64 |
| `convs.1.bn2.bias` | 64 |
| `convs.1.bn2.running_mean` | 64 |
| `convs.1.bn2.running_var` | 64 |
| `convs.1.bn2.num_batches_tracked` | 1 |
| `convs.2.fc_full.weight` | 21,632 |
| `convs.2.fc_full.bias` | 128 |
| `convs.2.bn1.weight` | 128 |
| `convs.2.bn1.bias` | 128 |
| `convs.2.bn1.running_mean` | 128 |
| `convs.2.bn1.running_var` | 128 |
| `convs.2.bn1.num_batches_tracked` | 1 |
| `convs.2.bn2.weight` | 64 |
| `convs.2.bn2.bias` | 64 |
| `convs.2.bn2.running_mean` | 64 |
| `convs.2.bn2.running_var` | 64 |
| `convs.2.bn2.num_batches_tracked` | 1 |
| `convs.3.fc_full.weight` | 21,632 |
| `convs.3.fc_full.bias` | 128 |
| `convs.3.bn1.weight` | 128 |
| `convs.3.bn1.bias` | 128 |
| `convs.3.bn1.running_mean` | 128 |
| `convs.3.bn1.running_var` | 128 |
| `convs.3.bn1.num_batches_tracked` | 1 |
| `convs.3.bn2.weight` | 64 |
| `convs.3.bn2.bias` | 64 |
| `convs.3.bn2.running_mean` | 64 |
| `convs.3.bn2.running_var` | 64 |
| `convs.3.bn2.num_batches_tracked` | 1 |
| `conv_to_fc.weight` | 2,048 |
| `conv_to_fc.bias` | 32 |
| `fc_out.weight` | 32 |
| `fc_out.bias` | 1 |

**Table S3.3. MEGNET hyperparameters.** The training of MEGNET models is based on its official repo v0.2.2 (https://github.com/materialsvirtuallab/megnet). The hyperparameters are recommended in the MEGNET paper and modified only with correspondence from the original authors.

| MEGNET Hyperparameters | Setting |
|---|---|
| MEGNET blocks | 3 |
| Minimum Epochs | 1000 (regression) / 500 (classification) |
| Initial atomic feature length | 95 |
| Element embedding length | 16 |
| Bond feature length | 100 |
| Hidden units in layer 1 | 64 |
| Hidden units in layer 2 | 32 |
| Hidden units in layer 3 | 16 |
| Batch size | 128 |
| Optimization algorithm | Adam |
| Learning rate | 0.001 (and auto-reduce if encountering nan) |
| Neighboring cutoff | 4.0 Angstrom |

**Table S3.4 MEGNet full model breakdown.** In similar fashion to Table S3.2, each layer and the number of trainable parameters is enumerated. The output shape of the layer and its connection to further layers within the overall architecture is also provided. In total, the model architecture specifies 167,761 trainable parameters. Layer types reference the syntax of the python neural network library Keras[17]; output shapes reference NumPy[18] NDArray representations.

| Layer (type) | Output Shape (numpy format) | Number of Parameters | Connected to |
|---|---|---|---|

| input_1 (InputLayer) | (None, None) | 0 | |
|---|---|---|---|
| embedding_1 (Embedding) | (None, None, 16) | 1,520 | input_1[0][0] |
| input_2 (InputLayer) | (None, None, 100) | 0 | |
| input_3 (InputLayer) | (None, None, 2) | 0 | |
| dense_1 (Dense) | (None, None, 64) | 1,088 | embedding_1[0][0] |
| dense_3 (Dense) | (None, None, 64) | 6,464 | input_2[0][0] |
| dense_5 (Dense) | (None, None, 64) | 192 | input_3[0][0] |
| dense_2 (Dense) | (None, None, 32) | 2,080 | dense_1[0][0] |
| dense_4 (Dense) | (None, None, 32) | 2,080 | dense_3[0][0] |
| dense_6 (Dense) | (None, None, 32) | 2,080 | dense_5[0][0] |
| input_4 (InputLayer) | (None, None) | 0 | |
| input_5 (InputLayer) | (None, None) | 0 | |
| input_6 (InputLayer) | (None, None) | 0 | |
| input_7 (InputLayer) | (None, None) | 0 | |
| meg_net_layer_1 (MEGNetLayer) | [(None, None, 32), …] | 39,392 | dense_2[0][0] dense_4[0][0] dense_6[0][0] input_4[0][0] input_5[0][0] input_6[0][0] input_7[0][0] |
| add_1 (Add) | (None, None, 32) | 0 | dense_2[0][0] meg_net_layer_1[0][0] |
| add_2 (Add) | (None, None, 32) | 0 | dense_4[0][0]<br>meg_net_layer_1[0][1] |
| add_3 (Add) | (None, None, 32) | 0 | dense_6[0][0]<br>meg_net_layer_1[0][2] |
| dense_7 (Dense) | (None, None, 64) | 2,112 | add_1[0][0] |
| dense_9 (Dense) | (None, None, 64) | 2,112 | add_2[0][0] |
| dense_11 (Dense) | (None, None, 64) | 2,112 | add_3[0][0] |
| dense_8 (Dense) | (None, None, 32) | 2,080 | dense_7[0][0] |
| dense_10 (Dense) | (None, None, 32) | 2,080 | dense_9[0][0] |
| dense_12 (Dense) | (None, None, 32) | 2,080 | dense_11[0][0] |
| meg_net_layer_2<br>(MEGNetLayer) | [(None, None, 32),<br>…] | 39,392 | dense_8[0][0]<br>dense_10[0][0]<br>dense_12[0][0]<br>input_4[0][0]<br>input_5[0][0]<br>input_6[0][0]<br>input_7[0][0] |
| add_4 (Add) | (None, None, 32) | 0 | add_1[0][0]<br>meg_net_layer_2[0][0] |
| add_5 (Add) | (None, None, 32) | 0 | add_2[0][0]<br>meg_net_layer_2[0][1] |
| add_6 (Add) | (None, None, 32) | 0 | add_3[0][0]<br>meg_net_layer_2[0][2] |
| dense_13 (Dense) | (None, None, 64) | 2,112 | add_4[0][0] |
| dense_15 (Dense) | (None, None, 64) | 2,112 | add_5[0][0] |

| Layer | Output Shape | Param # | Connected to |
|---|---|---|---|
| dense_17 (Dense) | (None, None, 64) | 2,112 | add_6[0][0] |
| dense_14 (Dense) | (None, None, 32) | 2,080 | dense_13[0][0] |
| dense_16 (Dense) | (None, None, 32) | 2,080 | dense_15[0][0] |
| dense_18 (Dense) | (None, None, 32) | 2,080 | dense_17[0][0] |
| meg_net_layer_3 (MEGNetLayer) | [(None, None, 32), …] | 39,392 | dense_14[0][0]<br>dense_16[0][0]<br>dense_18[0][0]<br>input_4[0][0]<br>input_5[0][0]<br>input_6[0][0]<br>input_7[0][0] |
| add_7 (Add) | (None, None, 32) | 0 | add_4[0][0]<br>meg_net_layer_3[0][0] |
| add_8 (Add) | (None, None, 32) | 0 | add_5[0][0]<br>meg_net_layer_3[0][1] |
| set2_set_1 (Set2Set) | (None, None, 32) | 2,640 | add_7[0][0]<br>input_6[0][0] |
| set2_set_2 (Set2Set) | (None, None, 32) | 2,640 | add_8[0][0]<br>input_7[0][0] |
| add_9 (Add) | (None, None, 32) | 0 | add_6[0][0]<br>meg_net_layer_3[0][2] |
| concatenate_1 (Concatenate) | (None, None, 96) | 0 | set2_set_1[0][0]<br>set2_set_2[0][0]<br>add_9[0][0] |
| dense_19 (Dense) | (None, None, 32) | 3,104 | concatenate_1[0][0] |
| dense_20 (Dense) | (None, None, 16) | 528 | dense_19[0][0] |
| dense_21 (Dense) | (None, None, 1) | 17 | dense_20[0][0] |

# 4. Automatminer Preset Comparison on Experimental Metallicity Classification

For tasks where the AutoML algorithm can fit and iterate models rapidly (*i.e.,* the dataset is small, <$10^4$ samples), Automatminer can require increasingly large computational effort to marginally improve predictive performance. Therefore, for small datasets, the bulk of Automatminer's performance can be retained using inexpensive presets (~1-5 minute training) versus more expensive presets (24h+ training).

Here we show the performance of three presets on the Matbench experimental metallicity classification task (sourced from Zhuo et. al[19]). Debug presets use only the most inexpensive yet information-dense featurizer (MagPie[4,10]), does only correlative feature reduction (based on Pearson correlation between sets of features), and is limited to 2 minutes AutoML training time. The Express preset uses slightly more expensive featurization and model-based feature reduction (as explained in Supplement Section 2 and in the main text) and uses a much longer maximum AutoML training time of 24 hours. Finally, the Heavy preset utilizes a wide range of Matminer featurizers, many of which are computationally expensive relative to the Express and Debug presets, and employs more expensive AutoML training with 48-hour time limit. Note the TPOT AutoML training algorithm can halt training early if the internal validation score does not improve over many training epochs; this is similar to "early stopping" of neural network training. The Express and Heavy presets marginally improve on the Debug ROC-AUC at the cost of *much* (more than 1000x) higher runtimes.

While this case is not representative of all supervised materials ML tasks even within the Matbench test suite, it illustrates that the gap between expensive and inexpensive pipelines may significantly narrow as the dataset size decreases.

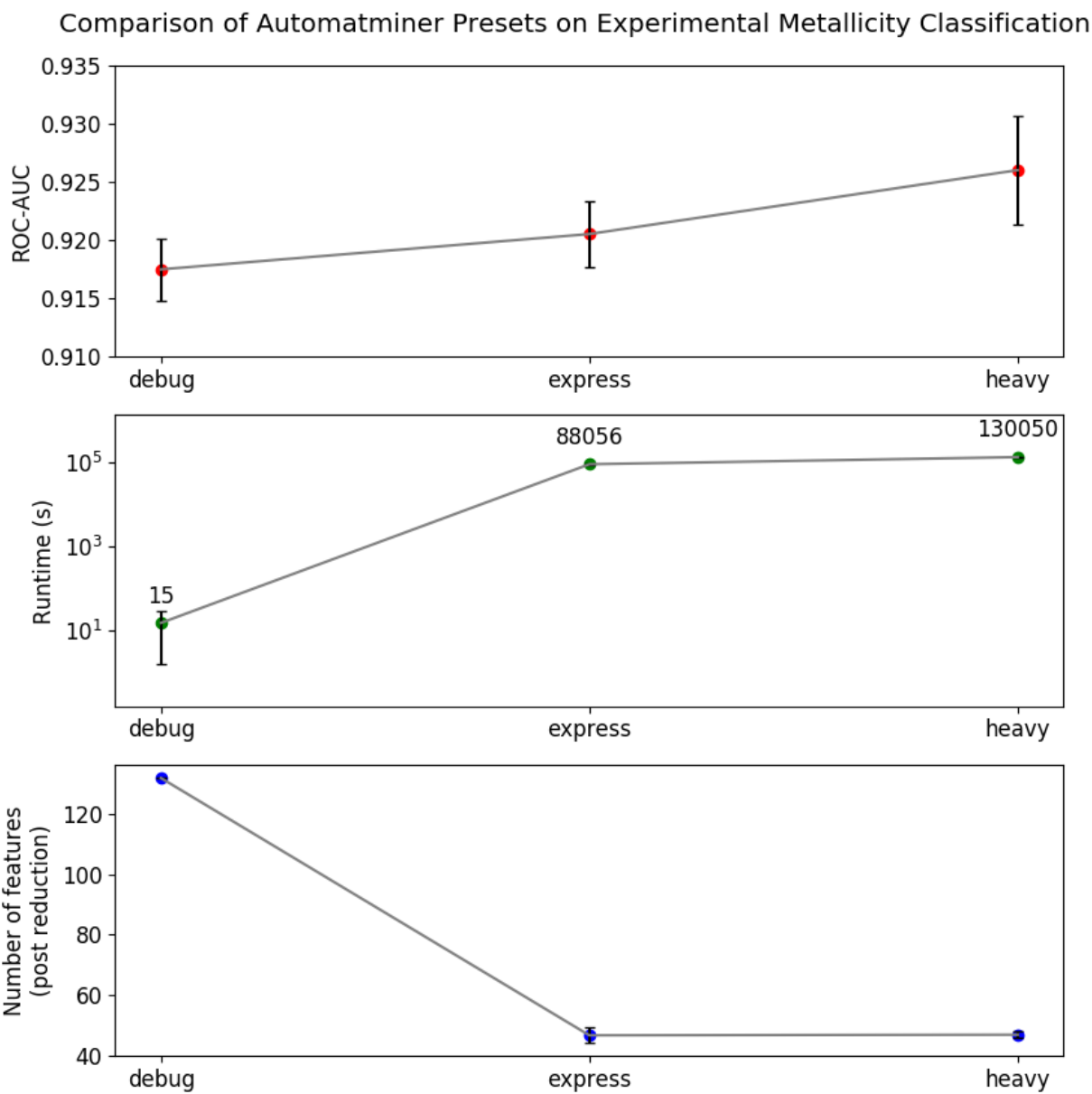

**Figure S2:** Comparison of Automatminer presets demonstrating diminishing AutoML returns on the Matbench experimental metallicity classification task. The abscissa shows three Automatminer presets – Debug, Express, and Heavy - in ascending order of computational intensity. Points represent the mean values among folds from a full 5-fold NCV evaluation on the metallicity task; error bars are the standard deviation between folds (error bars are ~<1% of mean value if not visible). ROC-AUC represents the receiver operating characteristic area under the curve for metallicity predictions; runtime is the elapsed time from the beginning of training to the end of prediction for each fold. The number of features is counted following feature reduction and represents the input to the AutoML algorithm. Debug has a much higher number of features because its preset undergoes minimal correlation-based feature reduction only.

# 5. Supplement References